\documentclass[prc,english,aps,twocolumn]{revtex4-1}
\usepackage[T1]{fontenc}
\usepackage[latin9]{inputenc}
\usepackage{amstext}
\usepackage{graphicx}
\usepackage{esint}

\makeatletter

\providecommand{\tabularnewline}{\\}

\@ifundefined{textcolor}{}
{%
 \definecolor{BLACK}{gray}{0}
 \definecolor{WHITE}{gray}{1}
 \definecolor{RED}{rgb}{1,0,0}
 \definecolor{GREEN}{rgb}{0,1,0}
 \definecolor{BLUE}{rgb}{0,0,1}
 \definecolor{CYAN}{cmyk}{1,0,0,0}
 \definecolor{MAGENTA}{cmyk}{0,1,0,0}
 \definecolor{YELLOW}{cmyk}{0,0,1,0}
}

\makeatother

\usepackage{babel}
\begin{document}

\preprint{This line only printed with preprint option}

\title{Fission properties of the BCPM functional}

\author{Samuel A. Giuliani}

\email{sam.and.giuliani@gmail.com}

\affiliation{Departamento de F\'\i sica Te\'orica, 
Universidad Aut\'onoma de Madrid, E-28049 Madrid, Spain}

\author{Luis M. Robledo}

\email{luis.robledo@uam.es}

\affiliation{Departamento de F\'\i sica Te\'orica, 
Universidad Aut\'onoma de Madrid, E-28049 Madrid, Spain}

\begin{abstract}
	
Fission dynamics properties of the Barcelona-Catania-Paris-Madrid 
(BCPM) energy density functional are explored with mean field 
techniques. Potential energy surfaces as well as collective inertias 
relevant in the fission process are computed for several nuclei 
where experimental data exists. Inner and outer barrier heights as 
well as fission isomer excitation energies are reproduced quite well 
in all the cases. The spontaneous fission half lives 
$t_{\textrm{sf}}$ are also computed using the standard semiclasical 
approach and the results are compared with the experimental data. 
The experimental trend with mass number is reasonably well 
reproduced over a range of 27 orders of magnitude. However, the 
theoretical predictions suffer from large uncertainties when the 
quantities that enter the spontaneous fission half life formula are 
varied. Modifications of a few per cent in the pairing correlation 
strengths strongly modify the collective inertias with a large 
impact  on  the spontaneous fission lifetimes in all the nuclei 
considered. Encouraged by the quite satisfactory description of the 
trend of fission properties with mass number we explore the fission 
properties of the even-even uranium isotope chain from $^{226}$U to 
$^{282}$U. Very large lifetimes are found beyond A=256 with a peak 
at neutron number N=184.

\end{abstract}

\maketitle


\section{Introduction}


Fission is a physical phenomenon taking place in heavy atomic nuclei 
that leads to the disintegration of a parent nucleus into two or 
more emerging fragments. It involves the evolution of the nucleus 
from its ground state to scission going through a variety of 
intrinsic shapes that cover a wide range of different intrinsic 
deformation parameters \cite{Krappe.12,Poenaru.96,Fission,Brack.72}. 
Fission properties depend upon the competition between the surface 
energy term coming from the strong nuclear interaction and the 
Coulomb repulsion and therefore they are often used as constraints 
and/or guidance to refine the parameters of effective nuclear 
interactions. A typical example is the D1S parametrizaton of the 
Gogny \cite{Decharge.80} force with parameters fine tuned to 
reproduce the fission barrier of $^{240}$Pu \cite{Berger.84}. More 
recently fission related constraints have been used with Skyrme 
interactions to define the UNEDF1 parametrization \cite
{UNEDF1,McDonnell.13}. 

The gross features of fission can be understood from a mean-field 
perspective using the Hartree-Fock-Bogoliubov (HFB) theory \cite
{Ring.80} and therefore it is not surprising the large amount of 
studies devoted to this subject with Skyrme  interactions \cite
{Bender.03,Erler.12,Staszczak.12,McDonnell.13}, Gogny ones \cite
{Berger.84,Egido.00,Warda.02,Delaroche.06,Dubrai.08,Martin.09,
Perez-Martin.09,Younnes.09} or based on the relativistic mean field \cite
{Abusara.10,Lu.12,Afanasjev.13}. Fission observables also depend on 
the inertia of the system to the relevant collective degrees of 
freedom and therefore they are sensitive to pairing correlations. As 
a consequence, fission is a good testing ground to test both the 
theories and interactions commonly used in nuclear structure. In 
addition, the theoretical understanding of fission is relevant to 
other areas outside traditional nuclear physics like safe energy 
production with nuclear reactors, radioactive waste degradation or 
the nucleosynthesis of heavy elements in the explosive galactic 
environments through the the r-process. Last but not least, a better 
understanding of fission could open the door to a better estimation 
of magic numbers and hence extra stability of super-heavy nuclei 
beyond Z=114. In this paper we explore the ability of a newly 
proposed energy density functional (EDF) denoted as 
Barcelona-Catania-Paris-Madrid (BCPM) \cite{BCPM} to describe 
fission. 

The BCPM  is a recent parametrization of the BCP EDF \cite
{Baldo.08,Baldo.10,Robledo.08,Robledo.10} devised for nuclear 
structure calculations. Its free parameters have been adjusted to 
reproduce the binding energies of even-even nuclei all over the 
nuclide chart, including deformed ones. Instead of the more 
traditional approaches where some central potential form is guessed 
(contact, gaussian, Yukawa, etc) and used afterwards to fit nuclear 
matter properties and/or the nuclear matter equations of state EoS 
(both symmetric and neutron), in the BCPM functional we start from a 
microscopic EoS that is fitted by means of a low order polynomial in 
the density. That polynomial fit is translated to a finite nuclei 
EDF just by replacing the nuclear matter density by the density of 
the finite nucleus. This procedure is inspired by the local density 
approximation (LDA) and is common practice in practical applications 
of the Kohn-Sham theory in condensed matter physics. The EDF is 
supplemented with a finite range surface term, a contact spin-orbit 
interaction of the same form as in Skyrme or Gogny forces, the 
Coulomb interaction and finally, a density dependent zero range 
pairing interaction \cite{Garrido.99} with strengths fitted to 
reproduce Gogny's neutron matter pairing gap. The parameters of the 
functional (essentially those of the finite range surface term plus 
some freedom in the polynomial fit to fine tune the binding energy 
per nucleon) are fitted to reproduce binding energies of the 518 
even-even nuclei of the 2003 mass table evaluation of Audi and 
Wapstra. The properties of the interaction concerning quadrupole, 
octupole and fission dynamics have also been explored \cite{BCPM}. 
As shown below, the BCPM functional gives reasonable results for 
fission observables including spontaneous fission half lives, 
fission isomer excitation energies, inner and outer barrier heights 
and mass distribution of fragments. We have also shown that those 
results could be improved by slightly modifying the amount of 
pairing correlations, either by modifying the pairing strengths or 
by going beyond the mean field approximation to restore the particle 
number symmetry broken by the HFB procedure. As a consequence of the 
satisfactory performance of BCPM in describing fission, we have 
explored fission properties of the uranium isotopic chain from 
proton drip line to the neutron drip line.

\section{Methods}

Both the BCPM energy density functional \cite{BCPM} and its 
predecesor BCP \cite{Baldo.08} are made of a bulk part which is 
determined by fully microscopic and realistic calculations of 
symmetric and neutron matter equations of state \cite
{Baldo.04,Taranto.13} as in the LDA of condensed matter physics. The 
two equations of state (symmetric and neutron matter) given as a 
function of the nuclear density are parametrized by low order 
polynomials of the densities. To account for finite size effects 
related to the surface energy, a phenomenological finite range 
gaussian interaction  is included. In addition, the Coulomb 
interaction and the spin-orbit term are taken exactly as in the 
Skyrme or Gogny forces. To deal with open-shell nuclei we include in 
the BCPM and BCP functionals a zero range density-dependent pairing 
interaction fitted to reproduce the nuclear matter gaps obtained 
with the Gogny force \cite{Garrido.99}. The calculations in finite 
nuclei are carried out with a modification of the HFBaxial computer 
code developed by one of the authors \cite{HFBaxial}.

To describe fission we follow the usual procedure based on the mean 
field approach with pairing correlations: the Hartree- Fock- 
Bogoliubov (HFB) theory with constraining fields. As constraining 
operators we have used mainly the axially symmetric quadrupole 
moment operator $Q_{20}=z^{2}-\frac{1}{2}(x^{2}+y^{2})$ although 
some exploratory calculations have also been performed with the 
octupole $Q_{30}$ and hexadecapole $Q_{40}$ moment operators and the necking operator 
$Q_{N}(z_{0})=\exp(-(z-z_{0})^{2}/C_{0}^{2})$. Axial symmetry is 
preserved in the calculations because of the high computational cost 
involved in the release of this restriction. We are aware of the 
relevance of triaxiality specially in the height of the inner 
fission barrier but its effect is merely quantitative and to a much 
lesser extent, qualitative. On the other hand, reflection symmetry 
is allowed to break at any stage of the calculation permitting 
octupole deformation and asymmetric fission. As a consequence of the 
breaking of the parity symmetry we are forced to constraint the 
center of mass to the origin as to prevent spurious solutions 
corresponding to a translation of the nucleus as a whole. The 
quasiparticle creation and annihilation operators of the HFB theory 
are expanded in a harmonic oscillator basis (HO) preserving axial 
symmetry and containing HO states with $J_{z}$ quantum numbers up to 
35/2 and up to 26 quanta in the $z$ direction. The basis contains 
over 3000 levels but time reversal invariance and the axial block 
structure reduces the computational complexity to a manageable 
level. The two lengths characterizing the HO basis, $b_{\perp}$ and 
$b_{z}$, have been optimized in a few nuclei for each value of the 
quadrupole moment. For the others the oscillator lengths computed 
for nearby nuclei are used. As the number of HFB configurations for 
each nucleus is large, a robust and fast gradient-like algorithm to 
solve the HFB equations is used \cite{Ring.80,rob11}. The most 
evident advantage of this method is the way it handles the 
constraints, which allows an easy generalization to an arbitrary 
number of them.

The spontaneous fission lifetime formulas Eqs \ref{eq:tsf} and \ref{eq:S}
below, depend crucially on the theory of the collective mass $B(Q_{20})$.
We shall use two methods to calculate it and compare in our results. The first
is the well-known cranking approximation to the Adiabatic Time dependent
HFB approximation \cite{Ring.80}. The resulting mass is expressed in terms of
the moments $M_{(-n)}$ of the generating field $Q_{20}$
\begin{equation}
	M_{(-n)} = \sum_{\alpha > \beta} \langle 0 | Q_{20} |\alpha \beta\rangle 
	\frac{1}{(E_{\alpha}+E_{\beta})^{n}} \langle \alpha \beta | Q_{20} |0\rangle 
\end{equation}
as
\begin{equation} \label{eq:ATDHFB}
	B(Q_{20}) = \frac{1}{2} \frac{M_{-3}}{(M_{-1})^{2}}.
\end{equation}
Here $|\alpha \beta\rangle$ are distinct 2-quasiparticles excitations and
$E_{\alpha}+E_{\beta}$ is the excitation energy, neglecting the quasiparticle-
quasiparticle interaction (cranking approximation \cite{Girod.79,Gia80,Libert.99}). 

An alternative method to calculate the mass is based on the gaussian overlap
approximation to the generator coordinate method (GCM). It is often simplified to obtain the expression 
\begin{equation}\label{eq:GCM}
	B(Q_{20}) = \frac{1}{2} \frac{M_{-2}^{2}}{(M_{-1})^{3}}.
\end{equation}

We shall calculate the lifetimes with both Eq \ref{eq:ATDHFB} and \ref{eq:GCM}
and compare. We note that Ref \cite{Baran.11} compares several forms of the mass, 
including Eq \ref{eq:ATDHFB} and \ref{eq:GCM}, in the context of the Skyrme
functionals. It is also important to mention the dependence of the mass with
the amount of pairing correlations: it has been shown in Refs \cite{Brack.72,Bertsch.91} that
the mass is inversely proportional to some power of the pairing gap. It means
that the stronger the pairing correlations are the smaller the mass is.

Zero point energy corrections to the HFB energy $\epsilon_{0}(Q_{20})$
are also considered in the ATDHFB and GCM approaches. In addition,
the rotational energy correction computed following the phenomenological
prescription of Ref \cite{egi04} is also substracted. This correction
is very important to the shape of the potential energy as its value
increases with deformation and can reach several MeV for large deformations.

The spontaneous fission half life is computed with the standard WKB
formalism of quantum mechanics. In the WKB formalism the $t_{{\rm sf}}$
is given (in seconds) by the formula 
\begin{equation} \label{eq:tsf}
t_{{\rm sf}}=2.86\,10^{-21}(1+\exp(2S))
\end{equation}
The action $S$ along the $Q_{20}$ constrained path is given by
\begin{equation} \label{eq:S}
S=\int_{a}^{b}dQ_{20}\sqrt{2B(Q_{20})(V(Q_{20})-(E_{\textrm{GS}}+E_{0)})}.
\end{equation}
where the integration limits correspond to the classical turning points
for the energy $E_{\textrm{GS}}+E_{0}$. For the collective quadrupole
inertia $B(Q_{20})$ we have used both the ATDHFB and the GCM expressions.
The results obtained with the two different theories can differ in
several orders of magnitude as the ATDHFB mass is known to be a factor
in between 1.5 and 2 larger than the GCM mass. The potential energy
$V(Q_{20})$ is given by the HFB mean field energy corrected by
zero point energies as described above, 
$V(Q_{20})=E_{\textrm{HFB}} (Q_{20})-\epsilon_{0} (Q_{20})-E_\textrm{Rot} (Q_{20})$.
 Finally, an additional parameter
$E_{0}$ is  added to the ground state
energy $E_{\textrm{GS }}$. It is meant to represent the  true ground
state energy obtained after considering quantal fluctuations in the
quadrupole degree of freedom. This quantity
could be estimated to be half of the square root of the curvature
around the minimum divided by the collective inertia but it is often
taken as a free parameter or kept fixed at some reasonable value.
We have followed the later approach with $E_{0}=1.0$ MeV and estimated
the impact of considering a larger value by repeating the calculations
with $E_{0}=1.5$ MeV. 

The spontaneous fission half life obtained in this way is subject to 
several uncertainties that can lead to differences of several orders 
of magnitude. The uncertainties are: 1) the height of the inner 
fission barrier gets reduced when triaxial shapes are allowed in the 
mean field calculation. The amount of reduction is typically of a 
couple of MeV, but it can show some isotopic variation (see \cite
{Delaroche.06} for a recent account in the actinide region); 2) the 
value of $E_{0}$. It can also make a difference in $t_{\textrm{sf}}$ 
specially for long lived isotopes where the fission barrier is wide. 
The reason is that the value of $E_{0}$ alters the classical turning 
points; 3) the values of the correlation energy corrections to the 
HFB energy included in $V(Q_{20})$. The values are computed under 
certain assumptions and approximations and a better estimation of 
their values can lead to some changes to $V(Q_{20})$; 4) the 
approximations involved in the evaluation of the collective masses 
can lead to differences of the order of 40 or 50 \%; 5) the  pairing 
correlation. It is an important ingredient both in the evaluation of 
the zero point energy  as well as in the evaluation of the 
collective inertia. As shown below, changes of a few percent in the 
pairing strength values can lead to changes in the theoretical 
estimation of the half lives in the range of five to twelve orders 
of magnitude. 

On the other hand, the experimental values of the parameters 
defining the potential energy of the fission process, namely the 
inner and outer barrier heights and the excitation energy of the 
fission isomer are  more robust quantities to compare with as they 
are not as sensitive to pairing correlations as the other 
parameters. However, these experimental quantities are obtained by 
model dependent assumptions that can mask the physical meaning  of 
the parameters. Therefore, although we have compared our values with 
the experimental ones, we prefer to compare the trends in 
spontaneous fission life times as a function of N and Z rather than 
the absolute values of the lifetimes themselves. We will also study 
the impact of changing various quantities entering the WKB formula. 

\begin{figure}[hbt]
\includegraphics[width=0.95\columnwidth]{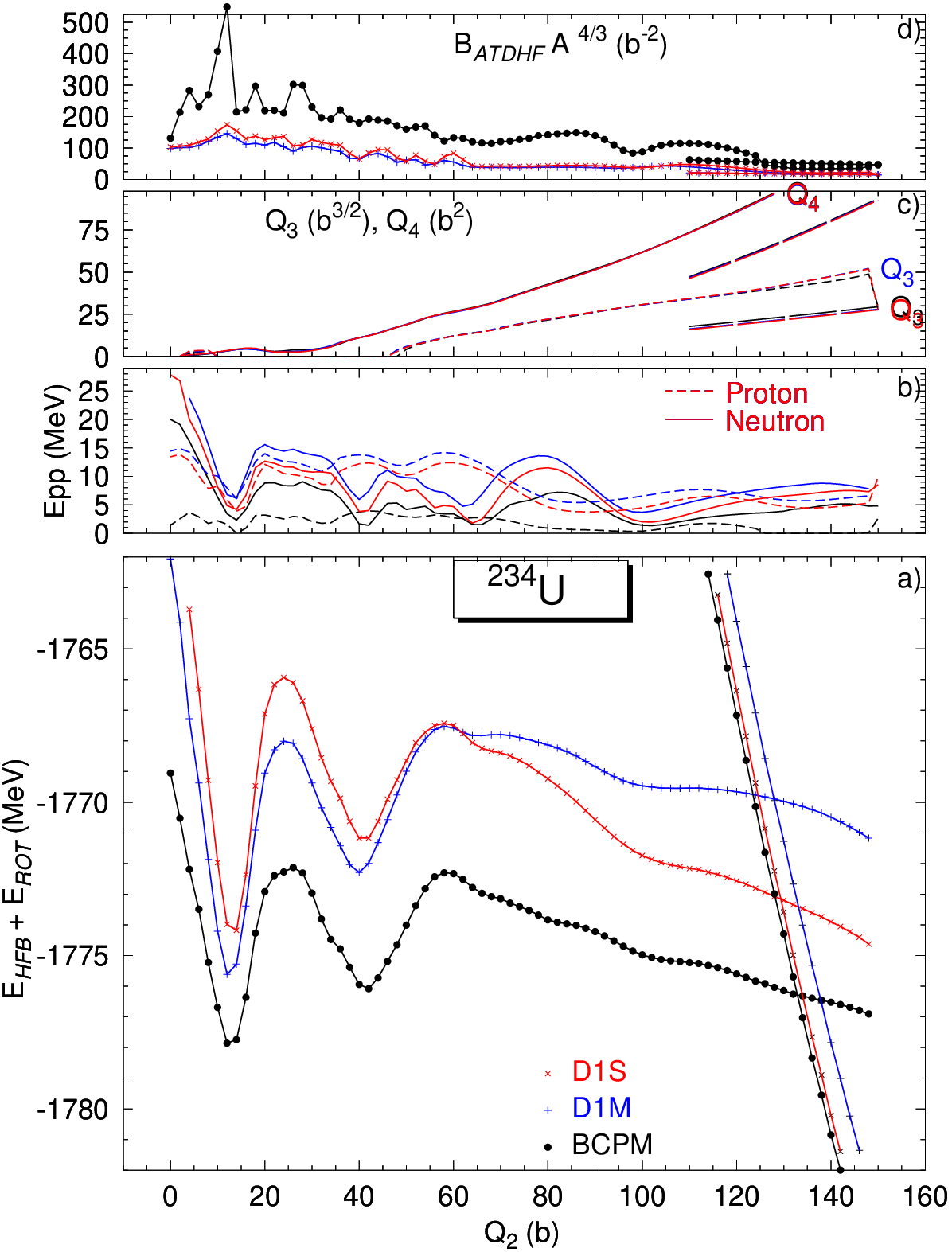}\caption{(Color online) 
Comparison of HFB mean field quantities as a function of the mass
quadrupole moment $Q_{20}$ obtained with three different interactions for the $^{234}$U nucleus.
The BCPM EDF (black curves, bullet symbol) as well as  the D1S (red, times symbol) and D1M 
(blue curves, plus symbol) parametrizations of the Gogny force are included.
In panel a) the HFB energy is given. In panel b)
the particle-particle correlation energy $E_{pp}=-\textrm{Tr}(\Delta\kappa)$
is plotted for protons (dashed lines) and neutrons (full lines) for
the three different kinds of calculations. In panel c) the
octupole and hexadecapole moments are given. Finally,
panel d) the ATDHFB collective inertia is depicted. \label{fig:Comparison}}
\end{figure}

\section{Results}


\subsection{Comparison with other interactions}


\begin{figure}[hbt]
\includegraphics[width=0.95\columnwidth]{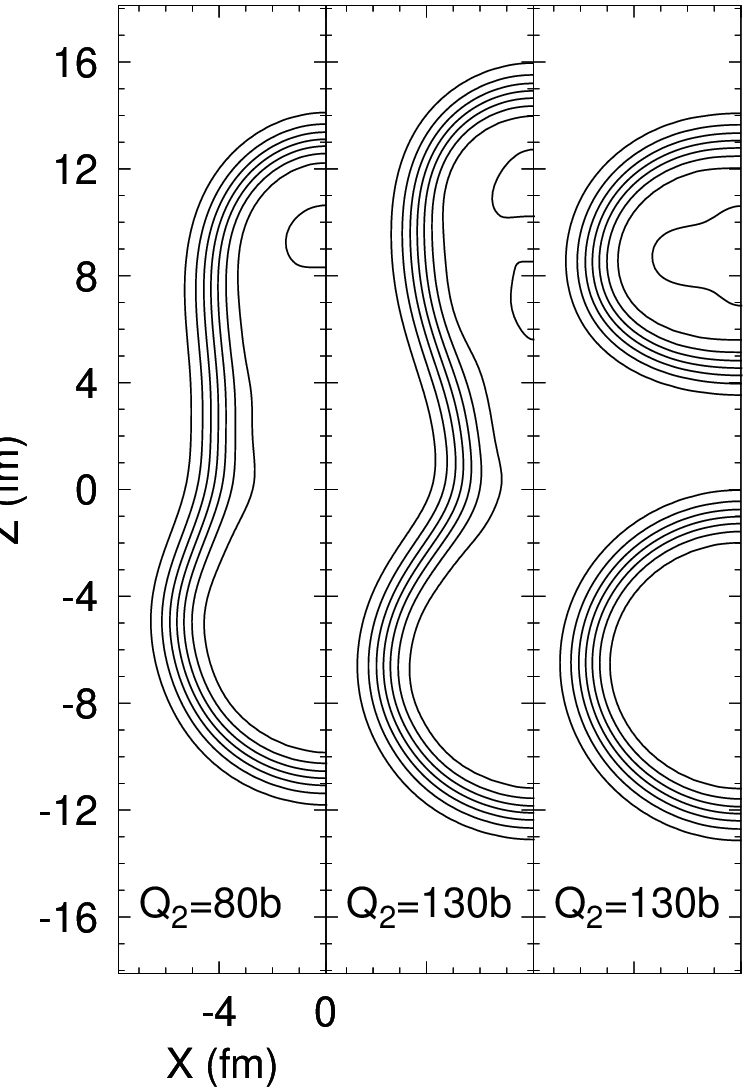}%
\caption{Contour plots of the densities of $^{234}$U for three 
different quadrupole deformation parameters indicated in each panel.
Contour levels correspond to densities between 0.02 fm$^{3}$ and 0.16 fm$^{3}$
in steps of 0.02 fm$^{3}$. The calculations correspond to the BCPM EDF. \label{fig:ContDens}}
\end{figure}

\begin{figure}[hbt]
\includegraphics[width=0.95\columnwidth]{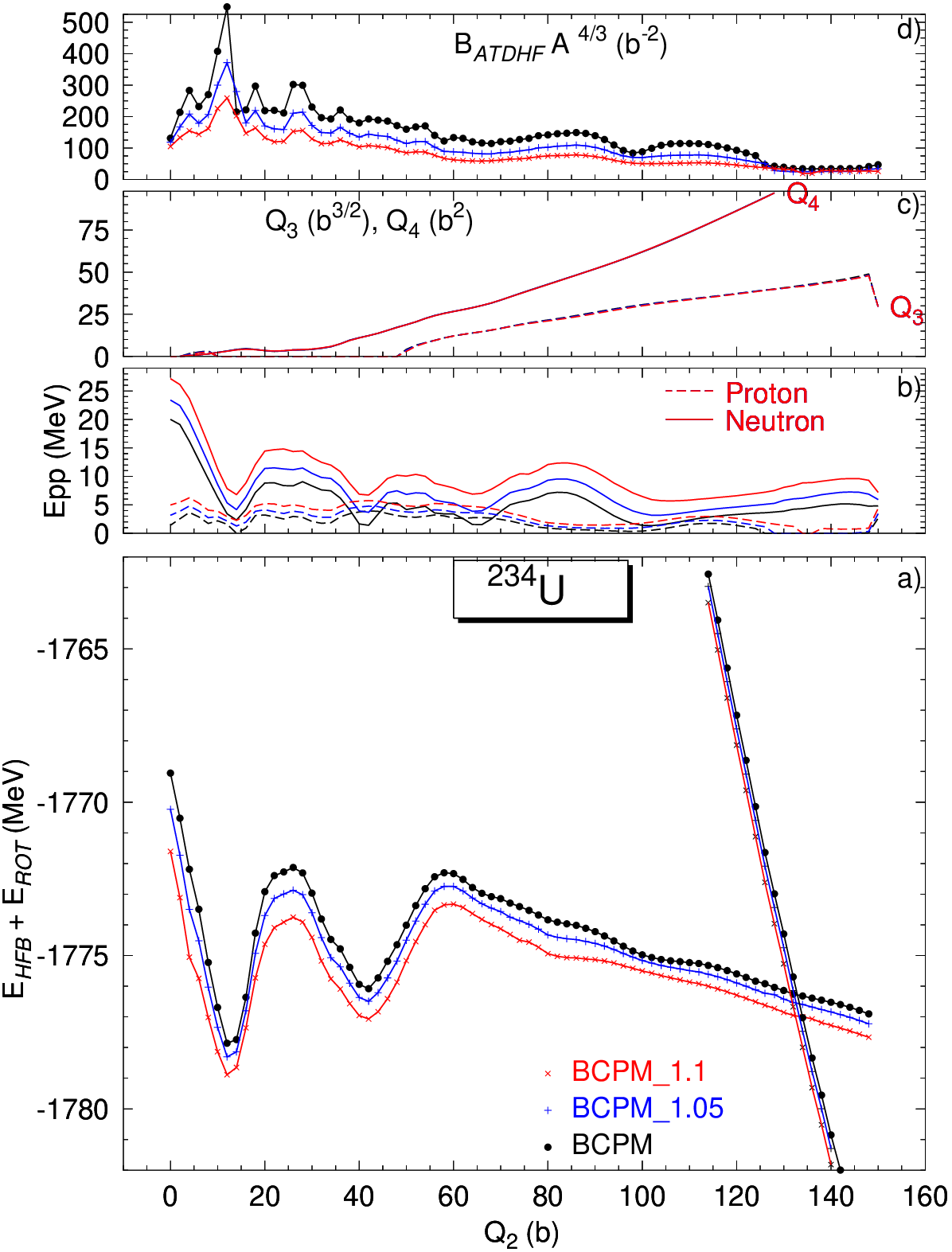}%
\caption{(Color online) Same as figure \ref{fig:Comparison} but for different pairing strengths.
The pairing strengths are given in terms of the reference value and
a scaling parameter $\eta$ taking the values 1.00 (the standard calculation),
1.05 and 1.10.\label{fig:Pairing}}
\end{figure}

Before comparing with the experimental data a comparison with the
HFB results obtained with the two Gogny interactions, namely, D1S
and D1M, is in order. The Gogny D1S interaction has been used in a
thorough study of heavy nuclei properties, including fission, in Ref
\cite{Delaroche.06} and it has proved to reproduce quite nicely most
of the properties analyzed. On the other hand, the fission properties
of D1M \cite{D1M} have not been analyzed in detail yet but its nice behavior regarding 
other aspects of nuclear structure like binding energies\cite{D1M}, radii \cite{RRG.10}
quadrupole \cite{Rob.08,RRG.10b} and  octupole \cite{Rob11b} properties make it a good candidate to compare
with. As we have already made comparison with D1S concerning fission
properties \cite{BCPM} of actinides ($^{240}$Pu) and super-heavies
($^{262}$Sg), we will just explore another actinide: the nucleus
$^{234}$U. 

In panel a) of Fig \ref{fig:Comparison} we compare the HFB energy as 
a function of $Q_{20}$ for the three functionals in the nucleus 
$^{234}$U. We observe how the shape of the three curves starting at 
$Q_{20}=0$b look rather similar up $Q_{20}=60$ b, apart from a 
constant shift of a few MeV. From there on, the D1M interaction 
declines more gently  than D1S. The BCPM results are in between D1M 
and D1S but closer to D1M than to D1S. The decline of energy with 
increasing quadrupole moment is correlated with the surface energy 
coeficient in nuclear matter which is larger in D1M and BCPM than in 
D1S \cite{BCPM}. The other three curves only present at large 
$Q_{20}$ values correspond to HFB solutions with two well separated 
fragments. They intersect the one fragment curves at $Q_{20}$ values 
around 130 b and they show a fast quasi-linear decrease in energy as 
the quadrupole moment increases. The charge and mass of each 
fragment are the ones that lead to the minimum energy for each 
quadrupole moment. For very large $Q_{20}$ values the distance 
between the fragments is larger than the range of the  nuclear 
strong interaction and therefore only the Coulomb repulsion energy 
between the fragments plays a role. In this two-fragment regime the 
quadrupole moment is directly linked to the separation distance 
between fragments \cite{Warda.11} and therefore increasing the 
quadrupole moment is equivalent to separating further away the 
fragments reducing the total energy of the system as a consequence 
of the reduction of Coulomb repulsion. Although the two-fragment 
curves seems to intersect the one-fragment ones this is just a 
consequence of projecting out paths in a multidimensional space of 
collective variables (quadrupole, octupole, hexadecapole, necking, 
etc) into an one dimensional plot (see below). There is a minimum 
action path with a ridge connecting both the one and the two 
fragments curves that goes along the multidimensional space. This 
path contributes to the action that enters the WKB formula  to 
compute the $t_\textrm{sf}$ half lives. As the determination of this 
path is cumbersome and its contribution to the action is small we 
will neglect its contribution to the action. This simplification 
amounts to consider both curves as really intersecting ones.  

A measure of pairing correlations, the pairing interaction energy 
$E_{pp}=-\textrm{Tr}(\Delta\kappa)$, is shown as a function of the 
mass quadrupole moment for protons (dashed lines) neutrons (full 
lines) in panel b) of Fig \ref{fig:Comparison}. Again the shape of 
the curves look rather similar for the three functionals but BCPM 
yields lower values of $E_{pp}$ than D1S and D1M. Those lower values 
correspond to less intense pairing correlations in the BCPM case 
with a severe quenching of the proton's pairing correlations. In  
panel c), the octupole and hexadecapole moments as a function of 
$Q_{20}$ are given. The results for the three EDFs are very similar, 
to the extent that they appear as a single curve for many $Q_{20}$ 
values. The full curves correspond to the one-fragment solutions 
whereas the long dashed ones correspond to the scissioned 
configurations. As mentioned before, the values of the multipole 
moments of the two kind of curves are quite different and therefore 
the paths are quite separated in the multidimensional space of 
parameters. Finally, in panel d) the collective inertia in the 
ATDHFB approximation is plotted. As in the case of the HFB and 
particle-particle energies the shape of the curves for the different 
interactions are quite similar but the BCPM inertia is up to a 
factor of three larger than the inertias obtained with the Gogny 
forces. The large value of the BCPM inertia is a direct consequence 
of the quenched pairing correlations: less pairing correlations 
imply a lower gap and therefore smaller two quasiparticle energies. 
As the two quasiparticle energies enter the collective inertia in 
the denominator, quenched pairing correlations imply enlarged 
collective inertias. On the other hand, the D1M inertia is around 15
\% smaller than the D1S one, consistent again with the quenched 
pairing correlations in D1M as compared to the D1S ones. The GCM 
inertias, not depicted, look rather similar in shape to the ATDHFB 
ones but are a factor of 0.5-0.6 smaller. The inertia for the two 
fragment solutions corresponds to the reduced mass of the two 
fragments \cite{Warda.11} and is constant with quadrupole deformation.

In the calculation of the $t_{\textrm{sf}}$ half life with the WKB 
formula the configuration with the lowest energy is always choosen. 
This is an approximation that neglects the path in the 
multidimensional space that connects the one fragment with the two 
fragment solution and therefore the $t_{\textrm{sf}}$ obtained are 
to be considered as lower bounds. On the other hand, triaxiallity 
leads to a reduction of the inner barrier height that is somehow 
compensated by an increase in the corresponding inertia \cite 
{Delaroche.06}. This effect has not been included in the present 
calculation. The values of the $t_{\textrm{sf}}$ half life obtained 
for the three EDFs computed with the GCM inertias are 
$t_{\textrm{sf}}=2.3\times10^{38}$s, $4.7\times10^{29}$s and 
$1.3\times10^{23}$s for BCPM, D1M and D1S, respectively. The large 
differences observed of up to 15 orders of magnitude can be 
attributed partly to the difference in the HFB energy curve but 
mostly to the very different values of the collective inertias. The 
previous $t_{\textrm{sf}}$ values have been obtained without taking 
into consideration the reduction of the inner barrier height as a 
consequence of triaxiallity. Also, increasing the value of the 
$E_{0}$ parameter to 1.5 MeV reduces the half lives by 6, 2 and 4 
orders of magnitude respectively. In any case, the values obtained 
are several orders of magnitude larger than the experimental value 
of $4.7\times 10^{23}$ s expect for Gogny D1S that is very close to 
experiment.  If the ATDHFB inertias are used instead of the GCM ones 
a much longer lifetimes are obtained: 
$t_{\textrm{sf}}=7.8\times10^{52}$s, $5.\times10^{40}$s and $2.9 
\times10^{32}$s for BCPM, D1M and D1S, respectively. This tendency 
to produce longer lifetimes when the ATDHFB inertias are used is 
common to all the isotopes considered in this study. The ATDHFB 
inertias are typically a factor 1.5 larger than the GCM ones (see 
Refs \cite{Baran.11,egi04} for examples) implying a 20 \%{} increase 
in the action and therefore a 20\% increase in the exponent of the 
lifetimes. This is a source of theoretical uncertainty in the 
evaluation of $t_{\textrm{sf}}$ that deserves further investigation. 
Another source of uncertainty comes from the fact that the inertias 
are computed in the "cranking approximation" where the energy of 
elementary excitations is replaced by the sum of HFB quasiparticle 
energies. The approximation \cite{egi04} can lead to overestimations 
in the inertias as large as 40 - 50\% for ground state 
configurations. Given the impact of these effects on the fission 
observables a better quantitative understanding is highly desirable.
In the following, to simplify the presentation, we will consider only
the GCM inertias in the evaluation of the lifetimes.

Finally in Fig \ref{fig:ContDens} contour plots of the densities for 
three values of the quadrupole moment are depicted. They are 
obtained in calculations with BCPM and differ little from the same 
quantities computed with D1M and D1S. For the quadrupole moment 
$Q_{20}$=130 b two densities, corresponding to the one fragment and 
two fragment solutions are presented. The two fragment solution 
shows a spherical fragment that corresponds to Z=51.60 and N=82.00 
and an oblate deformed fragment with Z=40.40 and N=60.00. The non 
integer proton and neutron number is due to the existence of low 
density nuclear matter between the two fragments. The oblate and 
slightly octupole deformed fragment ($\beta_{2}=-0.21$ and 
$\beta_{3}= 0.03$ acquires this shape to minimize the rather large 
Coulomb repulsion energy (assuming point like fragments the 
classical repulsion energy amounts to 196 MeV). This is an 
interesting result because it is commonly assumed \cite
{Moller.00,Moller.01} that the fission fragments can only have 
prolate shapes. 


\subsection{Varying pairing strengths}


\begin{table}
\begin{tabular}{ccccccc}
\hline 
Nucleus & $B_{I}^{\textrm{Th}}$ & $E_{II}^{\textrm{Th}}$  & $B_{II}^{\textrm{Th}}$ & $B_{I}^{\textrm{Exp}}$ & $E_{II}^{\textrm{Exp}}$ & $B_{II}^{\textrm{Exp}}$\tabularnewline
\hline\hline
$^{234}$U  & 5.87 & 1.78 & 5.59  &  4.80  &  --  & 5.50 \tabularnewline
\hline 
$^{236}$U  & 6.49 & 1.90 & 6.04  &  5.0   & 2.75 & 5.67 \tabularnewline
\hline  
$^{238}$U  & 6.99 & 2.03 & 6.54  &  6.30  & 2.55 & 5.50 \tabularnewline
\hline 
$^{238}$Pu & 6.91 & 1.85 & 5.20  &  5.60  & 2.4  & 5.10 \tabularnewline
\hline 
$^{240}$Pu & 7.43 & 2.08 & 5.69  &  6.05  & 2.8  & 5.15 \tabularnewline
\hline 
$^{242}$Pu & 7.72 & 2.27 & 6.30  &  5.85  & 2.2  & 5.05 \tabularnewline
\hline 
$^{244}$Pu & 7.89 & 2.47 & 6.30  &  5.70  &  --  & 4.85 \tabularnewline
\hline 
$^{240}$Cm & 6.8  & 1.2  & 3.90  &  --    & 2    & --   \tabularnewline
\hline 
$^{242}$Cm & 7.4  & 1.7  & 4.5   &  6.65  & 1.9  & 5.0  \tabularnewline
\hline 
$^{244}$Cm & 8.0  & 1.9  & 5.0   &  6.18  & 2.2  & 5.10 \tabularnewline
\hline 
$^{246}$Cm & 8.4  & 2.3  & 5.5   &  6.0   &  --  & 4.80 \tabularnewline
\hline 
$^{248}$Cm & 8.34 & 2.04 & 5.47  &  5.80  &  --  & 4.80 \tabularnewline
\hline 
$^{250}$Cf & 8.65 & 1.25 & 4.24  &  --    &  --  & 3.8  \tabularnewline
\hline 
$^{252}$Cf & 8.35 & 0.83 & 3.84  &  --    &  --  & 3.5  \tabularnewline
\hline 
\end{tabular}

\caption{Fission barrier height parameters $B_{I}$ (inner) and $B_{II}$(outer)
as well as excitation energy of the fission isomer $E_{II}$. The
three parameters are given in MeV. The theoretical values have been
obtained from the rotational energy corrected HFB potential energy
surface. The experimental values are taken from \cite{Sing.02} for
the $E_{II}$ and from \cite{Capote.09} for the $B$'s.\label{tab:Fission-barrier-heigth}}
\end{table}

In the BCPM functional the pairing interaction is taken as a density 
dependent contact pairing interaction with strength parameters fixed 
to reproduce the neutron matter pairing gap of the Gogny force \cite
{Garrido.99}. We have shown in the previous subsection that the 
particle-particle correlation energy, a quantity related to the 
amount of pairing correlations, was much smaller for BCPM than for 
the Gogny forces leading to much larger collective inertias. It is 
therefore reasonable to investigate the behavior of fission 
properties as a function of the pairing strength for the same 
functional. To this end, a parameter $\eta$ has been introduced as a 
multiplicative factor in front of the pairing gap field $\Delta_{kl}$
. For the sake of simplicity we have considered an unique parameter 
for both protons and neutrons although different parameters will 
give more flexibility to adjust the experimental data. The outcome 
of the calculations with $\eta$ values of 1.05 and 1.10 for the 
nucleus $^{234}$U are presented in Fig \ref{fig:Pairing}. We observe 
in panel a) that increasing the pairing strength by 10 \% (
$\eta=1.10)$ leads to an overall gain of the order of 1 MeV in 
binding energy. For the ground state the 1 MeV energy gains has to 
be compared to the 1 MeV of pairing correlation energy for the 
standard BCPM. The gain is even larger (1.6 MeV) for the 
configuration with $Q_{20}=26$ b and corresponding to the top of the 
inner barrier. However, the standard BCPM pairing correlation 
energy  is 2.06 MeV for that configuration. The net effect of 
increasing the pairing strength by 10\% is to  decrease the inner 
barrier eight ($B_{I})$ by 0.6 MeV whereas the other parameters, 
namely the outer barrier height $B_{II}$ and the fission isomer 
excitation energy remain more or less the same. The 
particle-particle correlation energies $E_{pp}$ shown in panel b) 
for protons and neutrons increase with increasing $\eta$ but the 
slope is larger for neutrons than for protons. The multipole moment 
values depicted in panel c) do not change at all when the pairing 
strenght is increased and the different values lie on top of each 
other for different $\eta$ values. Finally, the impact on the 
collective inertia is clearly visible in panel d) and is associated 
to the inverse dependence of the mass on the square of pairing gap 
\cite{Brack.72,Bertsch.91}. Increasing the pairing strength by 5 \% 
reduces the collective inertia by roughly 30 \% whereas a 10 \% 
increase leads to a reduction of 50 \%. The consequences of these 
reductions on the $t_{\textrm{sf}}$ are dramatic, decreasing its 
value by 11 orders of magnitude in going from $\eta=1.0$ to 
$\eta=1.05$ ($t_{\textrm{sf}}=8.0\times 10^{27}$ s) and six 
additional orders of magnitude in going from 
$\eta=1.05$ to $\eta=1.10$ ($t_{\textrm{sf}}=6.7\times 10^{21}$ s). This result is a clear indication of the 
very important role played by pairing correlations in the 
description of fission. The result suggests that experimental 
fission data could be used to fine tune the pairing strength instead 
of more traditional approaches based on odd-even staggerings. From a 
theoretical perspective the result also points to the very important 
role that the correlations associated to particle number symmetry 
restoration should have in fission dynamics. Restoring particle 
number symmetry usually leads to larger pairing correlations than 
the ones present at the mean field level and therefore will have a 
tremendous impact on fission lifetimes. In this respect it is worth 
mentioning that the dependence on density of BCPM is on integer 
powers of the density allowing the use of the regularization 
techniques suggested to solve some technical problems associated to 
the evaluation of the energy kernel overlaps required by symmetry 
restoration theories (see Refs \cite{Robledo.07,Robledo.10b} and 
references therein).

Given the large variability of the lifetimes with the different 
parameters entering the WKB formula, a direct comparison with the 
experimental data is meaningless and only comparisons with the 
trends along a series of nuclei or isotopes, all of them computed in 
the same conditions, can lead to meaningful conclusions regarding 
fission properties.


\subsection{Nuclei with known experimental data}


\begin{figure}[hbt]
\includegraphics[width=0.95\columnwidth]{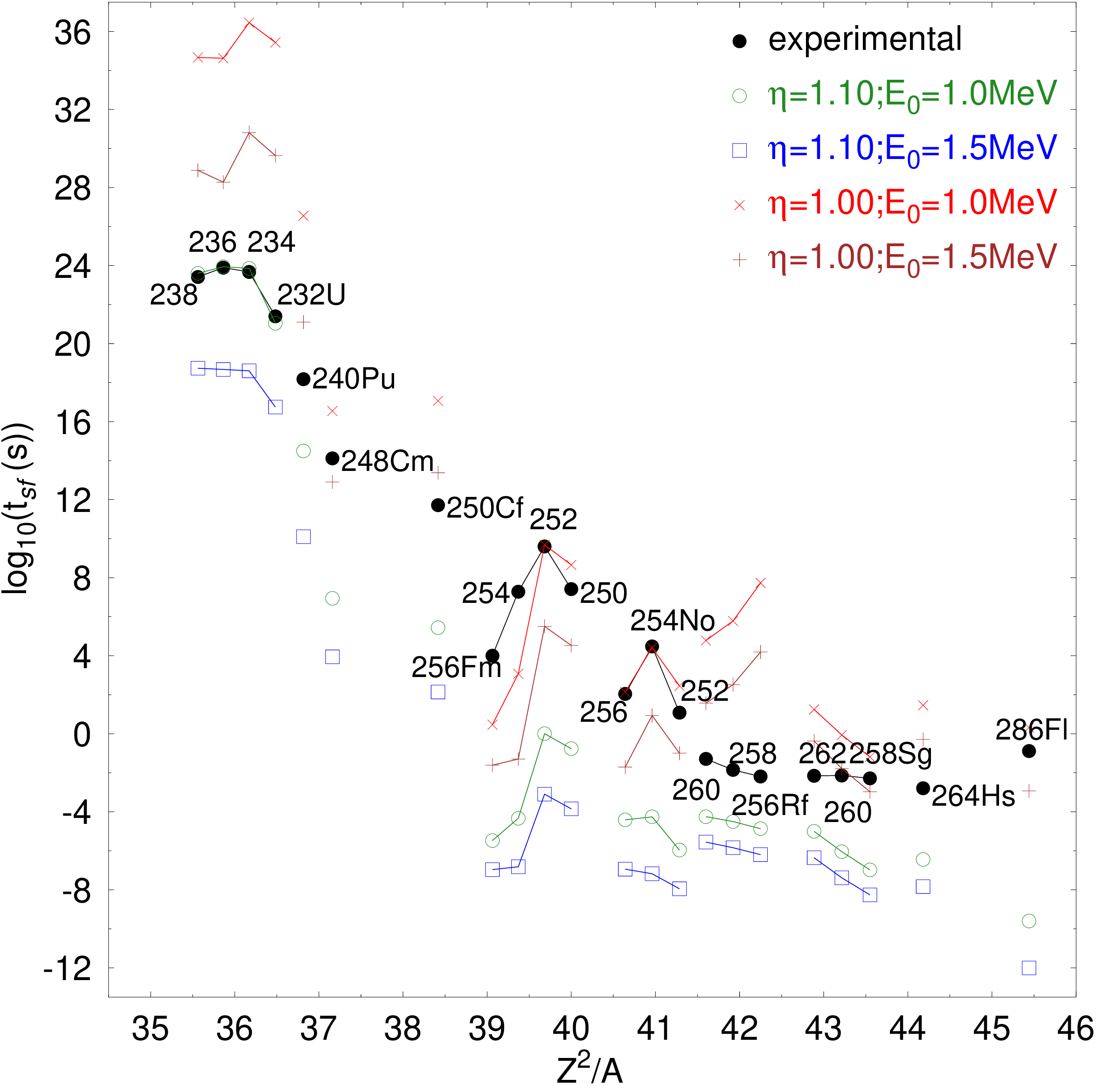}%
\caption{(Color online) Experimental $t_\textrm{sf}$ half lives (bullets) are 
compared to different theoretical results (open symbols) for several
isotopic chains where experimental data exists. The $t_\textrm{sf}$ 
are plotted as a function of the fissibility parameter $Z^{2}/A$.
See text for details. \label{fig:Spontaneus-fission-half}}
\end{figure}

\begin{figure}[hbt]
\includegraphics[width=0.95\columnwidth]{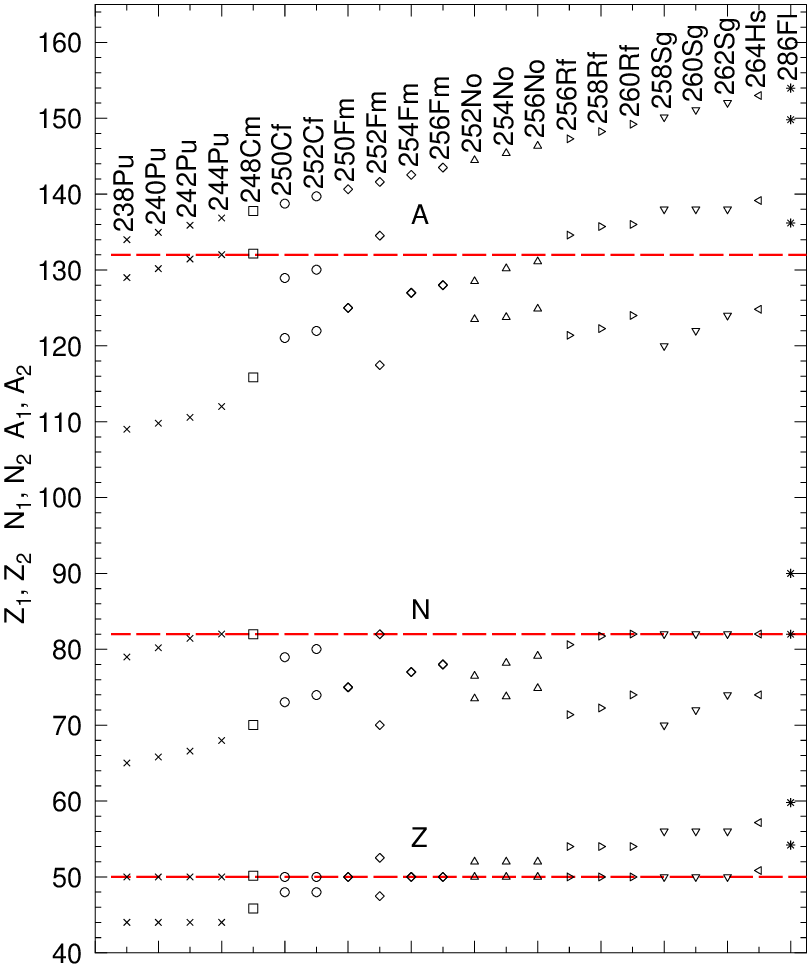}%
\caption{Proton ($Z_{1,2}$), neutron ($N_{1,2}$) and mass number ($A_{1,2}$) of the
two fragments emitted by the fission of the parent actinides (perpendicular
labels on top). The magic
numbers 50 and 82 as well as its sum are highlighted with horizontal lines.
\label{fig:FragMassDistAcinited}}
\end{figure}

In order to validate  BCPM as a functional able to describe fission properties, 
we have performed calculations for those even-even nuclei where the 
spontaneous fission half life has been measured. We will also 
compare the parameters defining the theoretical potential energy 
surface, namely the inner and outer barrier heights ($B_{I}$ and 
$B_{II}$) and the excitation energy of the fission isomer $E_{II}$ 
with available experimental data \cite{Sing.02,Capote.09}. It has to 
be mentioned that the experimental data for $B_{I}$ and $B_{II}$ 
\cite{Capote.09} is model dependent and therefore less reliable that 
the pure $t_{\textrm{sf}}$ data. In table \ref{tab:Fission-barrier-heigth} 
the experimental and theoretical values 
for $B_{I}$, $B_{II}$ and $E_{II}$ are given for all nuclei where 
experimental data exists \cite{Sing.02,Capote.09}. The theoretical 
values have been obtained by considering the HFB energy as a 
function of $Q_{20}$ with the rotational energy correction (computed 
in the way described in the previous section) subtracted. The effect 
of the zero point energy correction $\epsilon_{0}(Q_{20})$ has not 
been included mainly because it is almost constant as a function of 
$Q_{20}$. We notice that the theoretical predictions for $B_{I}$ are 
typically one or two MeV larger than experiment. This is not 
surprising as it is well known that the theoretical inner fission barrier is 
affected by triaxiallity and its height typically decreases by one 
or two MeV when the effect is included in the calculation \cite
{Delaroche.06}. Triaxiallity is not included at present because we 
still do not have access to a triaxial code incorporating the BCPM 
functional but work in this direction is in progress. The situation 
is slightly better in the comparison with the $E_{II}$ and $B_{II}$ 
values. For them, no significant triaxial effects are expected and the agreement 
with experiment is better than for the $B_{I}$. In Ref \cite{McDonnell.13} a 
thorough comparison of these data with various model predictions has 
been made. In this paper, the RMS deviations for the fission isomer energy and 
second barrier height are given for several mean field models. The 
BCPM values $\sigma (E_{II}) = 0.57$ MeV and $\sigma (B_{II})=0.72$ 
MeV are similar in magnitude to the ones of  UNEDF1 \cite{UNEDF1} a 
Skyrme variant specifically tailored to describe fission. This is a quite
satisfactory result taking into account that BCPM does not use any fission
data in its fit.

In Fig \ref{fig:Spontaneus-fission-half} the theoretical
$t_{\textrm{sf}}$ results obtained for different choices of the 
$E_{0}$ and $\eta$ parameters are compared to the known experimental 
values. The experimental $t_{\textrm{sf}}$ values \cite{holden.00} 
span a range of 27 orders of magnitude for a mass range A=232-286. 
The theoretical predictions, not including triaxial effects and 
computed with the GCM masses and zero point energies, span an even 
larger range of values and show a large variability depending upon 
the choices for the parameters. Focusing on the ``standard'' 
theoretical values $\eta=1.0$ and $E_{0}=1.0$ we observe differences 
with the experiment of up to 16 orders of magnitude for the lighter 
nuclei that steadily decrease to differences of just a couple of 
orders of magnitude for the heavier ones. The largest differences 
are observed for nuclei with higher and wider barriers where the 
impact of parameters like $E_{0}$ is larger. The comparison in 
isotopic chains indicate that the trend with neutron number 
compares much better with the experiment than the absolute values. 
The same conclusion can be extracted from the overall trend with 
mass number obtained from the table. Therefore, we conclude that the 
HFB predictions, although subject to large uncertainties due to 
uncontrolled approximations in the evaluation of the different 
parameters, can be used to guess with a reasonable precision the 
trends of $t_{\textrm{sf}}$ with mass number. The second conclusion 
drawn from this plot is the extreme sensitivity of the half lives to 
changes in $\eta$ and $E_{0}$: Increasing the pairing strength by 10\% ($eta=1.10$)
decreases $t_{\textrm{sf}}$ by several orders of magnitude. In the Uranium
isotopes the reduction is of 12 orders of magnitude bringing the theoretical
predictions on top of the experimental data. On the contrary, in the Fm and
No isotopic chains the reduction represents only 6 orders of magnitude but
worsens the agreement with experiment. On the other hand, the increase
of $E_{0}$ from 1 MeV to 1.5 MeV also reduces $t_{\textrm{sf}}$ by 
several orders of magnitude, but the reduction is not as severe as with
the increase of pairing strength. In the Uranium case, the reduction represents
on the average 6 orders of magnitude. Incidentally, the $t_{\textrm{sf}}$ values
obtained with $\eta=1$ and $E_{0}=1.5$ MeV are in most of the cases very
close to the results (not shown) corresponding to $\eta=1.05$ and $E_{0}=1$ MeV.

The sensitivity of the results to the pairing strength demands a theory
beyond HFB to describe pairing correlations. A first candidate would be
particle number restoration supplemented with configuration mixing using
the pairing gaps as collective coordinates. Also the sensitivity to the $E_{0}$
parameter justifies an effort to better understand its rationale. This is
obviously a task for the future.

\begin{figure}[htb]
\includegraphics[width=0.95\columnwidth]{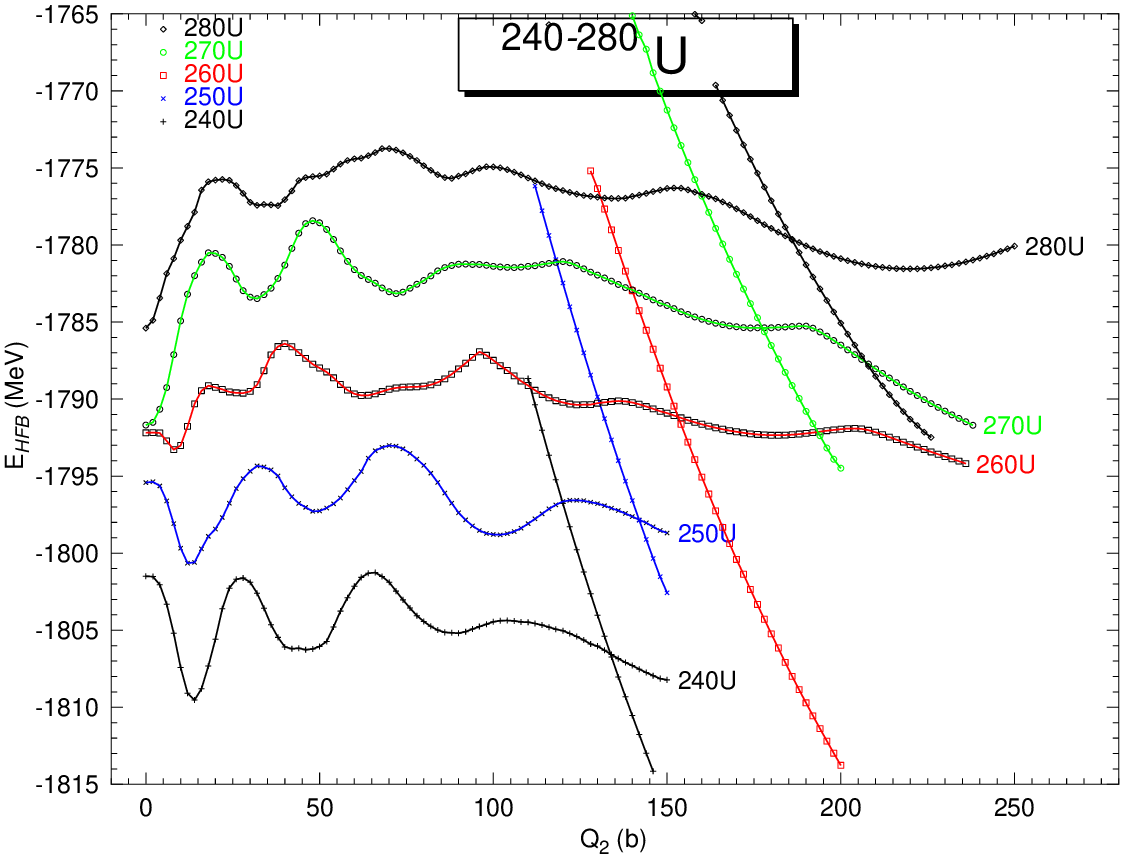}
\caption{(Color online) HFB energies as a function of the quadrupole moment $Q_{20}$ 
for some neutron rich uranium isotopes. The energies have been shifted 
upwards for the heavier isotopes in order to fit all the curves in a 
single plot. Along with the one fragment curves starting at $Q_{20}=0$ 
the curves corresponding to the energy of the two fragments resulting of
fission are given.\label{fig:HFB-Uranium}}
\end{figure}

Another important piece of relevant information is the mass 
distribution of the fission fragments. The mass of the fragments is 
determined by the nuclear shape in the neighborhood of  the scission 
point. As the scission point is difficult to characterize in a mean 
field theory that explores just a few degrees of freedom, we have 
preferred to take a different approach that involves the evaluation 
of quasi-fusion configurations. They are obtained by constraining 
the number of particles in the neck of the parent nucleus 
$$Q_{N}=\langle \phi| \hat{Q}_{N} (z_{0},C_{0}) | \phi \rangle $$ to 
a small value and then releasing the constraint to do a 
self-consistent calculation. Most of the time the self-consistent 
solution ends up in a solution with two well separated fragments. To 
make sure that the configuration  is the lowest in energy the 
procedure is repeated with different choices of the neck operator 
parameters $z_{0}$ (position of the neck along the $z$ direction) 
and $C_{0}$ (the width of the neck distribution). Those 
configurations are constrained to larger quadrupole moments in order to separate 
the fragments (please remember that for two fragments the quadrupole 
moment is proportional to the separation of the fragments \cite
{Warda.11}). An example of those quasi-fusion curves has already
been presented in Fig \ref{fig:Comparison}. In Fig \ref{fig:FragMassDistAcinited} the proton (Z) 
and neutron (N) numbers of the fragments obtained for the actinides 
considered are given. The Z values of the fragments are 
mostly determined by the Z=50 magic number except in $^{252}$Fm.  
Also for the heaviest nucleus considered $^{266}_{114}$Fl larger Z 
values are observed. For neutrons, the magic N=82 seems also to be 
dominant neutron number but here the exceptions are more numerous. For the 
plutonium and heavier isotopes the heaviest fragment has a mass 
number between 130 and 132, ten unit less than the average 
experimental value of Refs \cite{Hoffman.74,dematte.97}. The 
discrepancy can be attributed to the lack of quantum fluctuations in 
our model that can modify substantially the raw mean field numbers 
\cite{Goutte.05}.

Obviously, the numbers given here are meant to represent the peaks of
the fragments' mass distribution which is a broad distribution as a 
consequence of exchange of particles during the scission process as
well as a consequence of neutron evaporation. A better dynamical theory
is required (see \cite{Goutte.05} as an example of such theory) in order
to reproduce the experimental broad distribution.


\subsection{Neutron rich uranium isotopes}


In the previous section we concluded that the description of fission 
based on the HFB theory is subject to large uncertainties coming 
from the poor understanding of the way the different quantities 
entering the WKB formula should be computed. However, we also 
concluded that the HFB theory is reproducing reasonably well the 
experimental $t_{\textrm{sf}}$ trends with mass number. Encouraged 
by the result, we have performed calculations in the uranium 
isotopic chain from the light uranium $^{226}$U up to the neutron 
drip line corresponding to $^{282}$U with the aim of understanding 
and analyzing the trends in spontaneous fission half lives and the 
mass of the emerging fission fragments. To illustrate the results 
the HFB potential energies  
for the  $^{240}$U, $^{250}$U, $^{260}$U, $^{270}$U and $^{280}$U 
isotopes are depicted as a function of $Q_{20}$ in Fig \ref{fig:HFB-Uranium}. The energies of the heavier 
isotopes have been shifted by different amounts of energy (55, 100, 
135 and 160 MeV, respectively) to fit all the curves in a single 
plot. Also the curves corresponding to the two fragment solution 
with the lowest energy are depicted.

\begin{figure}[hbt]
\includegraphics[width=0.95\columnwidth]{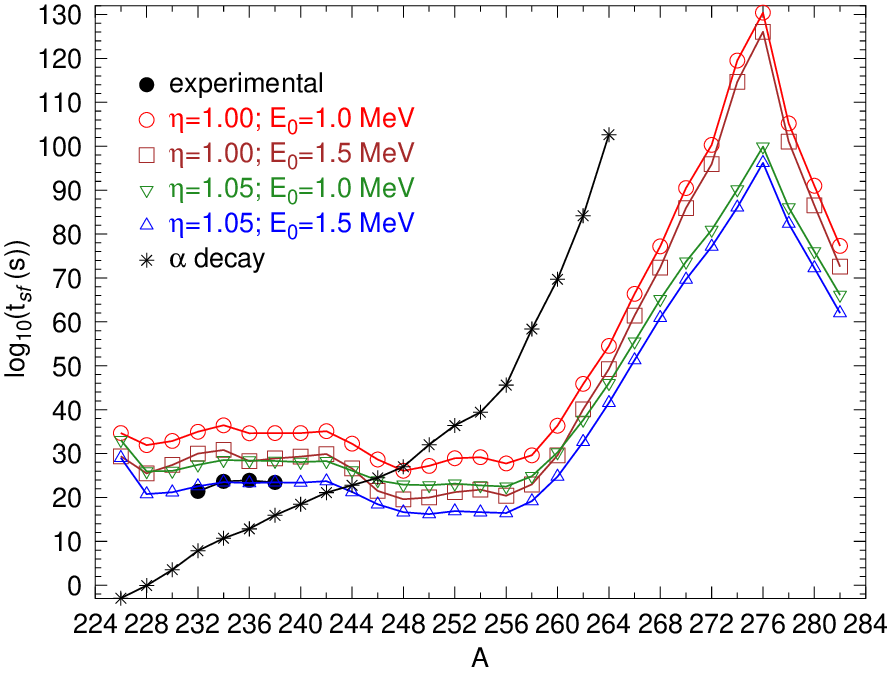}%
\caption{(Color online) Experimental $t_\textrm{sf}$ half lives (bullets) are 
compared to different theoretical results (open symbols, fission; asterisk $\alpha$-decay) for the
uranium isotopes up to the drip line nucleus $^{282}$U. \label{fig:tsfuranium}}
\end{figure}

We observe the ground state evolution from a quadrupole deformed 
ground state in $^{240}$U with $\beta_{2}=0.26$ to an spherical one 
for $^{270}$U (corresponding to N=178) and for $^{280}$U. It is 
also worth mentioning the existence of a second fission isomer in 
$^{240}$U (excitation energy of 3.9 MeV)  and $^{250}$U (excitation 
energy of 3.8 MeV, lower than the excitation energy of the first 
isomer). The situation in $^{260}$U is not as well defined as in the 
previous cases and there are three very swallow minima. The second 
one could be associated to the first fission isomer that is shifted 
to larger quadrupole moment values and zero octupole moment. Two 
fission isomers reappear in $^{270}$U but both are located at a very 
high excitation energy (around 9 MeV) and different quadrupole 
deformations than the ones in the light uranium isotopes. As a 
consequence of the increasing height and widening of the fission 
barriers as the neutron number approaches the neutron drip line we 
expect increasing $t_{\textrm{sf}}$ values as can be observed in the next
figure. In Fig 
\ref{fig:tsfuranium} the spontaneous fission half lives $t_{\textrm{sf}}$
of the uranium isotopes and computed with different choices of the $\eta${}
and $E_{0}$ parameters are plotted as a function of mass number A. 
As in previous cases, the $t_{\textrm{sf}}$ values have been 
obtained with the GCM collective mass and not taking into account 
the effects of triaxiallity in the first barrier. The 
usual range of up to 12 orders in magnitude depending on the choice of
parameters is observed. However, the trend with mass number A is the same
in all the four sets of parameters considered. This again gives us confidence
on the validity of the conclusions extracted from the trends with A. A decrease in the 
$t_{\textrm{sf}}$ values is observed up to mass number A=256 where
it becomes a steady increase with mass number up to  $^{276}$U where
the $t_{\textrm{sf}}$ values reach a maximum  that 
corresponds to a neutron number of 184 that corresponds to one 
of the classical magic numbers. The two neutron separation energy drops by
3 MeV in going from A=276 (4.99 MeV) to A=278 ($S_{2N}=1.94$ MeV what is
a clear indication of extra stability. The half lives for this and the other isotopes 
beyond $^{260}$U are very large and the corresponding nuclei can be 
considered as stable against the spontaneous fission decay channel. 
As the BCPM functional has been created to give a reasonable description
of masses, it is reasonable to use its predictions for the binding energies
of uranium and thorium to compute the half lives of $\alpha$-decay using
the phenomenological Viola-Seaborg formula \cite{Viola.66,Dong.05}. The 
results for the uranium isotopes are plotted as asterisks in the figure.
We observe a steady increase of $t_{\alpha}$ with mass number that reaches
its maximum at A=264 where $\alpha$ decay is no longer favorable energetically.
From A=244 on fission is faster than $\alpha$ decay.

\begin{figure}[hbt]
\includegraphics[width=0.95\columnwidth]{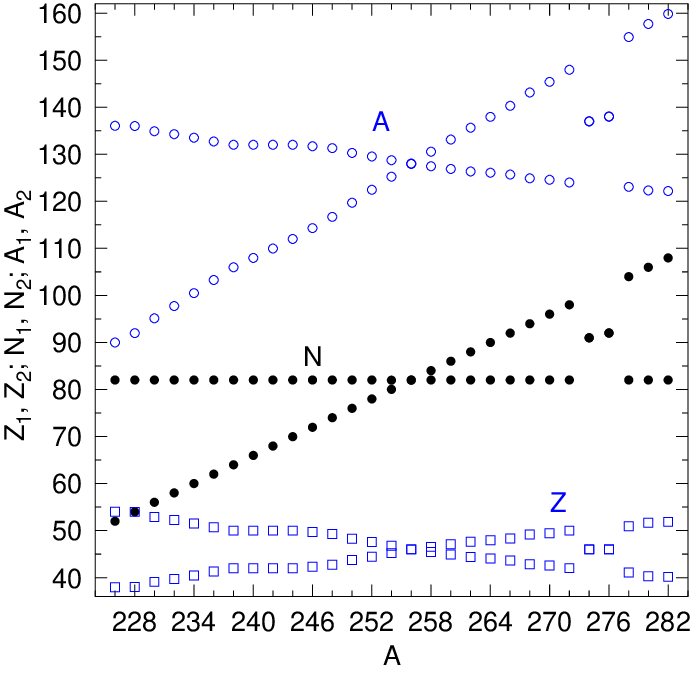}%
\caption{The number of protons ($Z_{1,2}$), neutrons ($N_{1,2}$) and
mass number ($A_{1,2}$) of the fission fragments is plotted as a function of the mass
number A of the parent uranium isotope.\label{fig:FragMassDistU}}
\end{figure}

In Fig \ref{fig:FragMassDistU} the proton, neutron and mass numbers 
of the fragments in the fission of the uranium isotopes are plotted 
as a function of mass number A. Those numbers are obtained by 
integrating the densities of each of the fragments coming out of the 
two fragment (fusion valley) self-consistent solutions mentioned 
before and corresponding to the lowest energy. In some cases, there 
are additional fusion valleys with different fragments but they lie 
higher in energy. The issue of how to describe dynamically the 
evolution of the system through those valleys is a very interesting 
subject of research (see for instance \cite{Goutte.05}) with many 
practical applications (the real fission fragment mass distribution) 
but exceeds the scope of the present paper. The numbers discussed 
below are to be taken as the ones of the peaks (most favorable mass) 
of the mass distribution of fragments prior to neutron emission. 
Except for $^{274}$U and  $^{276}$U the number of neutrons in one of 
the fragments always corresponds to the magic number N=82. The 
neutron number of the other fragment varies linearly accordingly to 
the mass number of the parent. On the other hand, the number of 
protons, that is close to the  magic number Z=50 for the light 
isotopes (in good agreement with experiment \cite{Schmidt.00}) 
varies linearly with mass number except for the isotopes $^{238-244}$
U where it stabilizes at Z=50 and for $^{274}$U and  $^{276}$U 
(symmetric fission) where it is distributed equally between the two 
fragments. Also for the $^{256}$U isotope a symmetric splitting with 
equal fragments is obtained. Concerning the mass distribution of the 
fragments, the heavy fragment has a mass number around 136 for the 
light isotopes that decreases with the mass number of the parent 
until the $^{256}$U isotope is reached. At this point the mass 
number of the heavy isotope starts to increase linearly with the 
exception of $^{274}$U and  $^{276}$U. This change in tendency is 
due to the increasing number of neutrons: as the number of neutrons 
increases, the fragment with N=82 is no longer the heaviest one. 
In any case, the behavior beyond A=256 is a rather academic issue 
as fission for those isotopes is a extremely unlikely process as 
discussed above.

\section{Conclusions}

The fission properties of several actinides and super-heavy nuclei 
have been computed with the recently proposed BCPM EDF. The 
theoretical results for the spontaneous fission half lives show a 
large variability consequence of uncertainties in the evaluation of 
some parameters of the theory and also on the strong dependence of 
the collective inertia with pairing correlations. As a consequence 
of the large uncertainties in the theoretical results we are only 
able to compare with the experimental data trend with mass number 
(for instance the reduction by 27 orders of magnitude in the 
spontaneous fission half lives in going from A=232 to A=286). 
The theoretical predictions seem to reproduce such trend giving us 
confidence in the convenience of the method and EDF for the study of fission
properties of neutron rich uranium isotopes. There we find that the 
spontaneous fission half lives remain more or less constant up to 
A=260 where they increase enormously as a consequence of the 
proximity to the magic neutron number N=184. Therefore, it is confirmed the prevalence
of this magic number in a extreme neutron rich case. On the other hand, a comparison of the parameters 
defining the potential energy surface for fission (inner and outer 
barrier heights and fission isomer excitation energies) with the 
model dependent ``experimental data'' show a rather good agreement 
that gives us additional confidence on the validity of our 
conclusions. The results obtained clearly show that more attention 
has to be paid to a proper description (including beyond mean field 
effects) of pairing correlations in the configurations relevant to 
fission. The evaluation of the quasi-fission valley in the HFB model
also allows to predict the peaks of the mass distribution of fission fragments.
It is shown that the magic proton number Z=50 and the magic neutron number
N=82 play an important role in determining the mass of the fragments.

To conclude, our study has shown the applicability of the BCPM EDF for fission 
studies in heavy and super-heavy nuclei. We have also pointed out
the large variability of the theoretical predictions to the models used
to evaluate the relevant parameters. However, this variability seems to respect
the trend with mass number of the spontaneous fission half life and therefore
we have applied our method to study fission properties of the uranium 
isotopes up to the neutron drip line. It is shown that beyond A=264
the uranium isotopes can be considered as stable against fission and $\alpha$-decay.

\begin{acknowledgments}
Work supported in part by MICINN grants Nos. FPA2012-34694, FIS2012-34479
and by the Consolider-Ingenio 2010 program MULTIDARK CSD2009-00064.
\end{acknowledgments}

\end{document}